\documentclass[pre,a4paper,onecolumn,notitlepage,nofootinbib]{revtex4-1}
\usepackage{graphicx}
\usepackage{amsmath}
\usepackage{amssymb}
\usepackage{bbm}
\usepackage{psfrag}
\usepackage{latexsym}
\begin{document}
\title{Free products of large random matrices - a short review of recent developments}

\author{Zdzislaw Burda}\email{zdzislaw.burda@uj.edu.pl}

\affiliation{Marian Smoluchowski Institute of Physics and 
Mark Kac Center for Complex Systems Research, Jagiellonian University, 
Reymonta 4, PL 30-059 Cracow, Poland}

\begin{abstract}
We review methods to calculate eigenvalue distributions of products of large random matrices. We discuss a generalization of the law of free multiplication to non-Hermitian matrices and give a couple of examples illustrating how to use these methods in practice. In particular we calculate eigenvalue densities of products of Gaussian Hermitian and non-Hermitian matrices
including combinations of GUE and Ginibre matrices. 
\end{abstract}

\maketitle

\section{Introduction}
Products of random matrices arise in many areas of research and engineering \cite{cpv}. In studies of dynamical systems one uses products of random matrices to derive limiting laws for Lyapunov exponents \cite{n}; in information theory and wireless telecommunication -- to calculate channel capacities for serial MIMO (multiple-input multiple-output) transmission \cite{m}. With the help of products of random matrices one investigates unitary evolution \cite{jw}, matrix diffusion \cite{gjjnw} or phase transitions in quantum Yang-Mills theories \cite{lnw}. One encounters products of random matrices in studies of quantum entanglement  \cite{cnz}, financial engineering \cite{blmp}, and many other fields of research \cite{bl,ckn}.

The problem of random matrix multiplication has attracted attention since the sixties \cite{fk, o}. Originally the studies were mostly concentrated on deriving limiting laws for products of infinitely many finite random matrices \cite{ckn} with some exceptions where also limiting laws for infinite matrices were formulated \cite{n}. Recently the focus has shifted to products of infinitely large matrices due to the discovery of a correspondence between infinitely large invariant random matrices and free random variables \cite{v0,s} which have been studied in the framework of free probability \cite{vdn} -- a new branch of probability theory that has been intensively developed since the eighties \cite{v1,v2}. Many results derived in this theory have been applied to large random matrices in the limit of infinite matrix size $N\rightarrow \infty$.
 
In this paper we review some of these results, and give practical recipes to calculate eigenvalue densities of products of large random matrices. Instead of giving formal proofs we gradually build up intuition passing from matrix addition through multiplication of Hermitian random matrices to multiplication of non-Hermitian matrices.  Free addition of large random matrices is a counterpart of addition of independent random variables in classical probability. We start with addition because it is much simpler than multiplication but it already has essential ingredients needed to understand a special role of invariant random matrices played in the formulation of limiting laws for eigenvalue distributions in the large $N$ limit. The key feature of invariant random matrices is that their eigenvectors are not correlated with the eigenvalues, and 
that they are uniformly distributed. The last statement means that the unitary matrix
constructed from the normalized eigenvectors is uniformly distributed on the unitary 
group. In effect, randomness of such matrices is entirely shaped by the probability distribution of the eigenvalues. The sum of independent invariant random matrices is an invariant random matrix.  In the limit $N\rightarrow \infty$ the eigenvalue density of the sum depends only on eigenvalue densities of the matrices added. This law is very similar to the law of addition in classical probability where the probability distribution of the sum of independent random variables can be calculated from the probability distributions of these variables. Similarities to classical probability go far beyond that. One can write a dictionary between classical and free probability. In this dictionary, random variable corresponds to invariant random matrix for $N\rightarrow \infty$, probability density function to eigenvalue density, characteristic function to Green's function, cumulant generating function (logarithm of the characteristic function) to R transform, Mellin transform of the probability density function to S transform, independence to freeness, etc. Using this correspondence one can find a full analogy between free addition and free multiplication of large invariant matrices and addition and multiplication of independent random variables.  

It is worth emphasizing that the laws of free addition and free multiplication can be independently derived using field theoretical methods, including perturbation theory of invariant matrix models \cite{bipz,biz}, without referring to free probability. Perturbation theory is a way of graphical enumeration of planar Feynman diagrams that arise in the large $N$ limit \cite{th}. A great advantage of the field theoretical approach is that it also works in the case of non-Hermitian matrices and it allows to derive a non-Hermitian version of the multiplication law \cite{bjn}. To the best of our knowledge this result is beyond the reach of free probability so far. We shortly recall this result here.

The paper is organized as follows. In section \ref{sec-add} we discuss eigenvalues of a sum of random matrices. In particular we define a slightly modified version of matrix addition -- such that it is independent of the eigenvectors of the matrices added. This brings us to the class of invariant random matrices that are discussed in section \ref{sec-inv}. In section \ref{sec-fadd} we recall the definition of the Green's function, the moment generating function and the R transform. These functions are used in the derivation of limiting laws for invariant matrices in the large $N$ limit. We employ these functions in the formulation of the law of free addition. In section \ref{sec-mult} we discuss the law of free multiplication for invariant Hermitian matrices. In section \ref{sec-iso} we discuss consequences of this law for multiplication of infinitely large isotropic non-Hermitian matrices. Finally in section \ref{sec-nh} we show how to calculate eigenvalue densities for products of large independent non-Hermitian matrices. We illustrate calculations with examples. 
We conclude the paper with a short summary in section \ref{sec-concl}.

\section{Eigenvalues of the sum of random Hermitian matrices}
\label{sec-add}

Let $a$ and $b$ be $N\times N$ Hermitian matrices.
We are interested in the eigenvalues of the sum $a+b$. To calculate them it is not sufficient to know the eigenvalues of $a$ and $b$. One would also need to know their eigenvectors. Actually it would be enough to know relative positions of the eigenvectors of $a$ and $b$: for example the decomposition of $b$-eigenvectors in the eigenbasis of $a$. One can, however, define a non-standard addition of matrices that depends only on the eigenvalues of individual matrices in the sum and not on their eigenvectors. This new addition is defined in a probabilistic way as a sum
\begin{equation} 
a \boxplus b = u^\dagger a u + v^\dagger b v \ ,
\label{fa}
\end{equation} 
where $u,v$ are independent Haar unitary matrices and $u^\dagger$ denotes the Hermitian conjugate of $u$. Let us recall that Haar unitary matrix is a random matrix distributed according to the uniform measure on the unitary group $U(N)$. The matrix $a \boxplus b$ is not a concrete matrix but it is a random matrix or equivalently an ensemble of matrices with a certain probability measure. The eigenvalues of $a \boxplus b$ are random variables. One is interested in the joint probability distribution of these random variables. The marginal distribution of the joint probability distribution is referred to as the eigenvalue distribution. One should note that the transformations $a \rightarrow u^\dagger a u$ and $b \rightarrow v^\dagger b v$ preserve the eigenvalues of the individual terms, but uniformly randomize their eigenvectors. One can therefore expect that the eigenvalue distribution of $a\boxplus b$ depends only on eigenvalues of $a$ and $b$. Moreover, since the sum  $u^\dagger a u + v^\dagger b v$ can be written as $u^\dagger(a + w^\dagger b w)u$, where $w=vu^{-1}$ is also a Haar unitary matrix, the eigenvalue distribution of the matrix $a+w^\dagger bw$ averaged over $w$ is the same as the eigenvalue density of $a \boxplus b$. In other words it is sufficient to randomize the eigenvectors of one matrix against the other.

One can consider a more general version of the problem in which $a$ and $b$ are independent random matrices instead of being deterministic ones. In this case the question is whether one can determine the eigenvalue distribution of $a\boxplus b$ from the eigenvalue distributions of individual matrices $a$ and $b$. The answer is affirmative for $N\rightarrow \infty$. In this limit the eigenvalue density $\rho_{a\boxplus b}(\lambda)$ of the sum $a\boxplus b$ can indeed be determined solely from the eigenvalue densities $\rho_a(\lambda)$ and $\rho_b(\lambda)$. For finite $N$ the answer is more complicated since it depends also on eigenvalue correlations. The correlations can be neglected only in the limit $N\rightarrow \infty$. The addition (\ref{fa}) is called free addition when $N\rightarrow \infty$ and  $a$ and $b$ are independent.  In the next section we shall see that the addition (\ref{fa}) is equivalent to the standard matrix addition $a+b$ for invariant random matrices.

\section{Invariant Hermitian matrices and free matrices}
\label{sec-inv}

Random matrices are formally defined by matrix ensembles equipped with probability measures. Let us denote the measure of an $N\times N$ Hermitian random matrix $x$ by $\mu(x)$. Random matrix $x$ is called invariant if the measure is invariant under the transformation $x \rightarrow U^\dagger x U$, where $U$ is an arbitrary unitary matrix: $\mu(x) = \mu(U^\dagger x U)$. Let us give two standard examples.

One can construct an invariant random matrix from a diagonal random matrix 
$\Lambda= {\rm diag}(\lambda_1,\ldots,\lambda_N)$ and Haar unitary random matrix $u$
\begin{equation}
a = u^\dagger \Lambda u \ .
\label{ulu}
\end{equation}
The invariance of the measure for $a$: $\mu(a)=\mu(U^\dagger a U)$ is a direct consequence of the invariance of the Haar measure $\mu_{Haar}(u)=\mu_{Haar}(uU)$ under the right multiplication by a unitary matrix $U$. One can choose the diagonal elements $\lambda_i$'s of $\Lambda$ to be independent, identically distributed real random variables. In such a case the joint eigenvalue density factorizes and $\rho_N(\lambda_1,\ldots,\lambda_N) = p(\lambda_1)\ldots p(\lambda_N)$ where $p(\lambda)$ is the probability density function for the random variables representing diagonal elements of $\Lambda$. The factorization of the joint eigenvalue density reflects the independence of the eigenvalues. The eigenvalue density is $\rho(\lambda)=p(\lambda)$.

In physical applications one frequently encounters invariant measures of the form \cite{mb,abf}
\begin{equation}
d\mu(a) \propto e^{ -N {\rm Tr} V(a)} da \ ,
\label{V}
\end{equation}
where $V(a)$ is a polynomial or a power series in $a$ and
$da = \prod_{1\le i\le N} 
da_{ii} \prod_{1\le i<j\le N} d {\rm Re} a_{ij} d {\rm Im} a_{ij}$, 
denotes flat measure for $N\times N$ Hermitian matrices. The proportionality sign in
(\ref{V}) means that the right hand side is displayed up to a constant. This implicit constant ensures the probabilistic normalization $\int d\mu(a) =1$.

The joint probability eigenvalue distribution is given by a well-known Dyson formula \cite{mb,d}
\begin{equation}
\rho_N(\lambda_1, \ldots, \lambda_N) \propto 
e^{-N\sum_i V(\lambda_i)} \prod_{i<j} (\lambda_j-\lambda_i)^2 \ ,
\label{rhoN}
\end{equation}
which can be interpreted as a partition function that describes a gas of $N$ particles in one dimension. We again skipped a normalization constant that ensures 
$\int_{R^N} \rho_N(\lambda_1, \ldots, \lambda_N) d\lambda_1 \ldots d\lambda_N = 1$.
In this picture the $i$-th eigenvalue $\lambda_i$ corresponds to the position of the $i$-th particle in one-dimensional space. Particles are in the potential $V(\lambda)$. They repel each other with a logarithmic potential depending of the distance $|\lambda_i - \lambda_j|$. The repulsion makes the eigenvalues interact. As a result they are correlated. The eigenvalue density can be calculated by integrating all but one eigenvalue from the joint eigenvalue distribution $\rho(\lambda)= \int\rho(\lambda,\lambda_2,\ldots,\lambda_N)  d\lambda_2\ldots d\lambda_N $. In the limit $N\rightarrow \infty$ 
one can relate the eigenvalue density to the potential by an integral equation \cite{bipz}
\begin{equation}
V'(x) = 2 \ {\rm P.V.} \int d\lambda \frac{\rho(\lambda)}{x-\lambda} \ 
\label{Vrho}
\end{equation}
obtained by applying the saddle point approximation to (\ref{rhoN}). P.V. stands for principal value. One can use this equation to calculate the eigenvalue density for the given potential or to find the potential corresponding to a given eigenvalue density. In particular one can choose the potential to reproduce the same eigenvalue density as in the first ensemble (\ref{ulu}) where the eigenvalue density is just equal to the probability density function for the diagonal elements of $\Lambda$. The two random matrix ensembles (\ref{ulu}) and (\ref{V}) have completely different statistics of eigenvalues even if they have identical eigenvalue densities.

Let us come back to the general discussion of invariant random matrices. The standard addition $a+b$ of invariant random matrices is equivalent to free addition $a\boxplus b$ (\ref{fa}). This follows from the invariance of the measure, which implies that the matrix $a$ is identically distributed as $u^\dagger a u$. The same holds for $b$ and $v^\dagger b v$. For $a$ and $b$ being invariant random matrices, the two sums $a+b$ and $a\boxplus b$ have identical probability measures. They represent the same random matrix ensemble. Therefore from here on we shall not distinguish between $a+b$ and $a\boxplus b$ while discussing invariant random matrices. The sum of invariant random matrices is also an invariant random matrix. An important property of the sum $a+b$ of invariant matrices is that the eigenvalue density depends in the limit $N\rightarrow \infty$ solely on the eigenvalue densities of $a$ and $b$, as follows from the discussion in section \ref{sec-add}. Independent invariant random matrices for $N\rightarrow \infty$ are called free. There is one-to-one correspondence between them and the free random variables known from free probability \cite{vdn}. 

To summarize this section, the standard addition of invariant random matrices is equivalent to free addition (\ref{fa}). In the limit $N\rightarrow \infty$ independent invariant random matrices are called free and the eigenvalue density $\rho_{a+b}(\lambda)$ of the free sum can be calculated from eigenvalues $\rho_a(\lambda)$ and $\rho_b(\lambda)$ of $a$ and $b$. The resulting density is independent of the detailed statistics of eigenvalues. For example, it does not depend on whether eigenvalues are correlated or not or whether they are generated using the ensemble of type (\ref{ulu}), (\ref{V}) or any other. The only thing that matters is the eigenvalue densities of the individual matrices in the sum. In the next section we briefly summarize the algorithm to calculate $\rho_{a+b}(\lambda)$. 

\section{Addition of free matrices}
\label{sec-fadd}

For $N\rightarrow \infty$ the eigenvalue density of an invariant random matrix $a$ can be described by an eigenvalue density $\rho_a(\lambda)$ having support $I$ on the real axis. Typically $I$ is a finite interval. In calculations it is convenient to use the Stieltjes transform of the density
\begin{equation}
G_a(z) = \int_I \frac{\rho_a(\lambda) d\lambda}{z-\lambda} \ ,
\label{green}
\end{equation}
that is a complex function defined on the complex plane outside $I$.
This function is often called Green's function. The density can be reconstructed from the Green's function $G_a(z)$ by calculating the imaginary part of the Green's function close to the real axis 
\begin{equation}
\rho_a(\lambda) = -\frac{1}\pi \lim_{\epsilon \rightarrow 0^+} 
{\rm Im} G_a(\lambda +i \epsilon) \ .
\label{rhoG}
\end{equation} 
The power expansion $G_a(z)$ at infinity 
\begin{equation}
G_a(z) = \sum_{n=0}^\infty \frac{m_{a,n}}{z^{n+1}}
\label{gexp}
\end{equation}
generates moments of the eigenvalue distribution
\begin{equation}
m_{a,n} = \int_I d\lambda \rho_a(\lambda) \lambda^n \ .
\end{equation}
These moments are equivalent to the trace moments 
\begin{equation}
m_{a,n} = \left\langle \frac{1}{N} {\rm Tr} a^n \right\rangle 
= \int d\mu(a) \frac{1}{N} {\rm Tr} a^n ,
\end{equation}
calculated with respect to the probability measure $\mu(a)$ for the entire matrix $a$. The zeroth moment is fixed $m_{a,0}=1$ by the normalization of the measure. One often defines another moment generating function as a power series at zero
\begin{equation}
\phi_a(z) = \sum_{n=1}^\infty m_{a,n} z^n \ . 
\label{phi}
\end{equation} 
It is related to the Green's function 
\begin{equation}
\phi_a(z) = \frac{1}{z} G_a\left(\frac{1}{z}\right) - 1 \ .
\label{phiG}
\end{equation}
The eigenvalue density $\rho_a(\lambda)$, the moment generating function $\phi_a(z)$, and the Green's function $G_a(z)$ contain roughly speaking the same information about the eigenvalue distribution of the invariant matrix $a$ and one can reproduce these functions from each other. The idea is to calculate these functions for $a+b$ when the corresponding functions for $a$ and $b$ are given. Let us first calculate the moments of $a+b$
\begin{equation}
m_{a+b,n} = \left\langle \frac{1}{N} {\rm Tr} (a+b)^n \right\rangle = 
\int \int d \mu_a(a) d\mu_b(b) \frac{1}{N} {\rm Tr} (a+b)^n \ .
\label{mab}
\end{equation}
The measure factorizes, since $a$ and $b$ are independent by assumption.
For $n=1,2,\ldots$ the last equation gives 
\begin{equation}
\begin{split}
m_{a+b,1} & = m_{a,1} + m_{b,1} \ , \\ 
m_{a+b,2} & = m_{a,2} + m_{b,2} + 2m_{a,1} m_{b,1} \ , \\
  & \ldots \ .
\label{mab2}
\end{split}
\end{equation}
In the second equation we replaced $\left\langle \frac{1}{N} {\rm Tr} ab \right\rangle$ by $m_{a,1} m_{b,1}$. The equality $\left\langle \frac{1}{N} {\rm Tr} ab \right\rangle=m_{a,1} m_{b,1}$ follows from the equation $\int\int d\mu_a(a) d\mu_b(b) \frac{1}{N} {\rm Tr} (a- \mathbbm{1} m_{a,1}) (b- \mathbbm{1} m_{b,1})=0$, which holds in the limit $N\rightarrow \infty$ as a result of invariance and independence of $a$ and $b$. The relations between moments (\ref{mab2}) can be written in a compact way using free cumulants which are certain (specific) combinations of moments \cite{vdn}: $\kappa_{a,1}=m_{a,1}$, $\kappa_{a,2} = m_{a,2} - m^2_{a,1}$, $\kappa_{a,3}=m_{a,3}-3m_{a,2}m_{a,1} + 2m_{a,1}^3$, etc. Equations (\ref{mab2}) correspond to 
\begin{equation}
\kappa_{a+b,n}= \kappa_{a,n} + \kappa_{b,n} \ , 
\label{kab}
\end{equation}
for $n=1,2,\ldots$. The concept of free cumulants is analogous to the standard cumulants in classical probability. Let us recall that cumulants for a real random variable are generated as coefficients of the power series expansion of the logarithm of the characteristic function. The logarithm of the characteristic function for a sum of independent random variables is equal to the sum of logarithms of the characteristic functions of these random variables. Therefore cumulants for the sum of independent variables are additive, exactly as free cumulants (\ref{kab}). Free cumulants are, however, generated in a different way. The generating function for free cumulants is called R transform \cite{vdn,v1}. Free cumulants are equal to coefficients of the power series expansion
\begin{equation}
R(z) = \sum_{n=1}^\infty \kappa_n z^{n-1} \ .
\label{R}
\end{equation}
Note that there is a shift between the power of $z$ and the index of the corresponding cumulant.
The R transform is related to the Green's function which generates moments of the eigenvalue distribution. The relation reads
\begin{equation}
G(z) = \frac{1}{z-R(G(z))} \ .
\label{gr}
\end{equation} 
We give it without a proof \cite{vdn}. One can find a diagrammatic interpretation of this equation \cite{bjn} using field theoretical methods for planar graphs enumeration \cite{bipz,biz}. Using equation (\ref{gr}) one can find $G(z)$ if $R(z)$ is known. One can also invert equation (\ref{gr}) 
\begin{equation}
G\left(R(z) + \frac{1}{z}\right) = z \ .
\label{rg}
\end{equation}
This form is useful when one wants to determine the R transform for
the given Green's function. The R transform for the sum of free matrices $a+b$ is a sum of 
R transforms for $a$ and $b$ \cite{vdn,v1}
\begin{equation}
R_{a+b}(z) = R_a(z) + R_b(z) \ ,
\label{RR}
\end{equation}
exactly as the logarithm of the characteristic function for the sum of independent real
random variables in classical probability. The additivity of R transform reflects the invariance and independence of the measures for $a$ and $b$ in (\ref{mab}). 
For $N\rightarrow \infty$ these two properties correspond to freeness.

Explicit relations between free cumulants $\kappa_n$ and moments $m_n$ can be found
by inserting the power series  $R(z) = \kappa_1  + \kappa_2 z + \ldots$ 
and $G(z)=1/z + m_1/z^2 + \ldots$ into equation (\ref{rg}). The first three free cumulants  are identical as the standard ones $\kappa_{1}=m_{1}$, $\kappa_{2} = m_{2} - m^2_{1}$, $\kappa_{3}=m_{3}-3m_{2}m_{1} + 2m_{1}^3$. From the fourth one on they are different. The fourth free cumulant is $\kappa_4 = m_4 - 4m_1 m_3 - 2 m_2^2 + 10 m_1^2 m_2 - 5 m_1^4$
while the standard one is $\kappa_4= m_4 - 4m_1 m_3 - 3m_2^2 + 12 m_1^2 m_2 - 6m_1^4$.

The R transform was first introduced in free probability by Voiculescu \cite{v1}. Later it was observed that free cumulants are related to a combinatorial problem of non-crossing lines on a plane \cite{s,ns1}. For this reason free cumulants are sometimes called non-crossing cumulants. For completeness we mention that equation (\ref{rg}) can be deduced \cite{bjn,z} using field theoretical methods of planar graph enumeration \cite{bipz}.  The Green's function and the R transform correspond to sums over certain classes of diagrams, and equations like (\ref{gr}) - to Dyson-Schwinger equations \cite{bipz}. The combinatorics for planar Feynman diagrams for matrix models is equivalent to the combinatorial approach to freeness \cite{s,ns1}. 

We close this section by giving a standard example. Recall that in classical probability the normal random variable with the probability density function $p(x) = \frac{1}{\sqrt{2\pi}} e^{-x^2/2}$ has all cumulants equal zero except $\kappa_2=1$. 
What is the corresponding random matrix eigenvalue density in free probability that 
has all free cumulants equal zero except the second one $\kappa_2=1$? We want to 
find the eigenvalue density corresponding to the R transform equal $R(z)=z$ (\ref{R}). 
Using (\ref{gr}) we first find an equation for the Green's function $G(z)= 1/(z-G(z))$. 
It can be solved for $G(z)$: $G(z)=\frac{1}{2} (z-\sqrt{z^2-4} )$. 
We select the branch of the square root to match the asymptotic 
behavior $G(z) \sim 1/z$ for large $z$. The eigenvalue density is (\ref{rhoG}) 
$\rho(\lambda) = \frac{1}{2\pi}\sqrt{4-\lambda^2}$ for $\lambda \in [-2,2]$ 
and zero otherwise. It is the Wigner semicircle law. Odd moments of this 
distribution are equal zero while even moments $m_{2k}=\int_{-2}^{2} d\lambda \rho(\lambda) \lambda^{2k} = \frac{1}{k+1}\binom{2k}{k}$ 
are equal to the Catalan numbers. One can use equation (\ref{Vrho}) 
to find that the corresponding potential is quadratic $V(x)=x^2/2$, 
so the invariant measure (\ref{V}) is Gaussian $d\mu(a) \propto e^{-N {\rm Tr} a^2/2}$. Let us slightly modify the R transform and consider $R(z) =  \alpha + \sigma z$. 
Without repeating the calculations one can easily argue on general grounds that the corresponding eigenvalue density is shifted as compared to the previous case by 
the first cumulant $\kappa_1=\alpha$ and rescaled by the second one $\kappa_2=\sigma$ leading to $\rho(\lambda)=\frac{1}{2\pi \sigma}\sqrt{4 -(\lambda-\alpha)^2/\sigma^2}$ 
and $V(x)=(x-\alpha)^2/2 \sigma^2$. 

Let us now calculate the eigenvalue density of the sum of two independent large ($N\rightarrow \infty$) invariant matrices $a$ and $b$ that have eigenvalue densities given by the Wigner semicircle law with different shift and scale parameters. Denote these parameters by $\alpha_a$, $\sigma_a$ for $a$ and $\alpha_b$, $\sigma_b$ for $b$, respectively. The R transforms are $R_a(z) = \alpha_a + \sigma_a z$ and $R_b(z) = \alpha_b+\sigma_b z$. 
To find the eigenvalue density of the sum $c=a+b$ we first calculate R transform that is equal to $R_c(z) = \alpha_c + \sigma_c z$ with $\alpha_c=\alpha_a+\alpha_b$ and $\sigma_c=\sigma_a+\sigma_b$. The R transform is exactly of the same form as for $a$ and $b$. Therefore the eigenvalue density for $c$ is also given by the Wigner semicircle law. The situation is analogous to that in classical probability where the sum of two random Gaussian variables is known to be a Gaussian random variable. One says that the semicircle law is stable with respect to free addition in the same way as the Gaussian law is stable with respect to addition in classical probability. Actually there exist a bijection between stable laws in classical and free probability that allows one to classify all stable laws in free probability \cite{bp}. Just as a curiosity we mention that the only distribution that is stable with respect to addition (random variables) and  free addition (random matrices) is the Cauchy distribution.
  
\section{Products of large invariant matrices}
\label{sec-mult}

As we have seen the law of addition for independent large invariant matrices can be concisely expressed in terms of the R transform (\ref{RR}). Before discussing the law of multiplication it is convenient to recall the corresponding law in classical probability. Let $x$ and $y$ be independent non-negative real random variables. What is the distribution of the random variable $xy$? A way to derive this distribution is via the Mellin transform of the probability density function. The Mellin transform $M_{xy}(s)$ for the product $xy$ is the product of Mellin transforms $M_{xy}(s)=M_x(s) M_y(s)$ for $x$ and $y$. Clearly this multiplication law is analogous to the addition law discussed in the previous section (\ref{RR}).  It remains to find the corresponding law for products of large invariant matrices. Let $a$ and $b$ be independent invariant matrices. For the moment we also assume $a$ and $b$ to be positive semi-definite. Later we will weaken this restriction. The first trivial observation is that even if $a$ and $b$ are Hermitian, the product $ab$ is generically non-Hermitian, so its eigenvalues are complex. Therefore, it is convenient to introduce a slightly modified version of the multiplication \cite{rs}
\begin{equation} 
a\cdot b = \sqrt{a} \; b \; \sqrt{a} \ .
\label{cprod}
\end{equation}
We use dot in the notation to distinguish $a \cdot b$ from the standard matrix multiplication $ab$. Square root of a positive semi-definite matrix is defined as  
$\sqrt{a} = U^\dagger {\rm diag}(\sqrt{\lambda_1},\ldots,\sqrt{\lambda_N})U$ where
$U$ is a unitary matrix that diagonalizes $a$: 
$a = U^\dagger {\rm diag}(\lambda_1,\ldots,\lambda_N)U$.
The product $a \cdot b$ of Hermitian matrices is Hermitian. 
The product $a \cdot b$ of invariant random matrices is an invariant random matrix. Moreover, the probability measure $\mu_{a \cdot b}$ for the product $a \cdot b$ is identical as for $b \cdot a$: $\mu_{a \cdot b} = \mu_{b \cdot a}$ so this product is in a sense commutative. One should also note that the trace moments for the product
$a\cdot b$ (\ref{cprod}) are identical as for the standard product: 
$\frac{1}{N} {\rm Tr} (a\cdot b)^k = \frac{1}{N} {\rm Tr} (ab)^k$.

The law of free multiplication was formulated by Voiculescu in free probability \cite{vdn,v2}. Later it was shown that it applies also to multiplication (\ref{cprod}) of invariant random matrices in the limit $N\rightarrow \infty$ \cite{rs}. The multiplication law is expressed in terms of S transform
\begin{equation}
S_{a\cdot b}(z) = S_a(z) S_b(z)
\label{Sab}
\end{equation}
where $S_a(z)$ is a complex function. The S transform $S_a(z)$ can be constructed from the moment generating function $\phi_a(z)$ (\ref{phi}) or actually the inverse function of $\phi_a$ that is usually denoted by $\chi_a$: $\chi_a(\phi_a(z))=\phi_a(\chi_a(z))=z$. The relation reads \cite{v2}
\begin{equation}
\label{S_chi}
S_a(z) = \frac{z+1}{z} \chi_a(z) \ .
\end{equation} 
The S transform can be viewed as a series in $z$
\begin{equation}
S_a(z) = \frac{1}{m_{a,1}} + \frac{m_{a,1}^2-m_{a,2}}{m_{a,1}^3} z + 
\frac{2 m_{a,2}^2 - m_{a,2}m_{a,1}^2 - m_{a,1} m_{a,3}}{m_{a,1}^5} z^2 + \ldots
\end{equation}
that is obtained by inserting the inverse series in place of $\phi_a(z)$ in equation (\ref{S_chi}). The coefficients of this series are related to the moments $m_{a,k}$. The coefficients are singular for $m_{a,1}=0$, in which case the S transform is ill-defined. 

The S transform is related to the R transform as follows \cite{vdn,bjn}
\begin{equation}
S(z) = \frac{1}{R(z S(z))} \ .
\end{equation}
If one knows the R transform then the last equation can be used to calculate the S transform. One can invert the last equation \cite{bjn}
\begin{equation}
R(z) = \frac{1}{S(z R(z))} \ ,
\end{equation}
which allows one to calculate the R transform if the S transform is known. Using these relations one can show that the multiplication law (\ref{Sab}) can be alternatively written in the R transform form \cite{bjn}:
\begin{equation}
\begin{split}
R_{a\cdot b}(z) & = R_a(w) R_b(v) \ ,\\
v & = z R_a(w) \ ,\\
w & = z R_b(v) \  .
\label{Rprod}
\end{split}
\end{equation}
This set of equations looks more complicated than (\ref{Sab}) but in some situations it is more advantageous. For example, it can be used when the first moments vanish, that is when the S transforms are ill-defined. Moreover, this form can be generalized to the case of non-Hermitian matrices, as we discuss in section \ref{sec-nh}. The two forms of the multiplication law are in a sense complementary. The original S transform form of (\ref{Sab}) has a very simple and appealing structure that reflects important properties of free products, for example that the  multiplication $a \cdot b$ (\ref{cprod}) of invariant random matrices is commutative and associative: $S_{(a\cdot b) \cdot c}(z) = S_{c \cdot (a \cdot b)}(z)$. This follows from the fact that multiplication on the right hand side of (\ref{Sab}) is just multiplication of complex numbers which is commutative and associative. 

Let us give an example to illustrate how these formal relations work in practice. Consider the product $a \cdot b$ of two independent identical Wishart matrices \cite{wish}. The Wishart matrix can be thought of as a square of a Gaussian matrix from the GUE ensemble \cite{mb}. The matrices $a$ and $b$ have the eigenvalue densities $\rho_a(\lambda) = \rho_b(\lambda) = \rho(\lambda)$
\begin{equation}
\rho(\lambda)=\frac{1}{2\pi} \sqrt{\frac{4-\lambda}{\lambda}}
\label{Wish}
\end{equation}
for $\lambda \in [0,4]$ and $\rho(\lambda)=0$ for $\lambda$ outside this interval.
The Green's function (\ref{green}) corresponding to this density is 
\begin{equation}
G(z) = \frac{1}{2} - \frac{1}{2} \sqrt{\frac{z-4}{z}}
\end{equation} 
and the moment-generating function (\ref{phiG}) 
\begin{equation}
\phi(z) = \frac{1}{2z}\left(1 - \sqrt{1-4z}\right) - 1 \ .
\end{equation}
The inverse function of $\phi$ fulfills the equation
$z = \frac{1}{2\chi(z)}\left(1 - \sqrt{1-4\chi(z)}\right) - 1$ which can be solved for $\chi(z)$
\begin{equation}
\chi(z) = \frac{z}{(1+z)^2} \ .
\end{equation}
Thus the S transform is (\ref{S_chi})
\begin{equation}
S(z) = \frac{1}{1+z} \ .
\end{equation}
We now want to find the S transform for the product $S_{a\cdot b}(z)$ from $S_a(z)=S_b(z)=S(z)$. Using the multiplication law (\ref{Sab}) we obtain
\begin{equation}
S_{a\cdot b}(z) = \frac{1}{(1+z)^2} \ .
\end{equation}
We now proceed to find the eigenvalue density $\rho_{a\cdot b}(\lambda)$ that corresponds to the S transform given above. First we write the equation for the inverse function of the moment generating function $\chi_{a \cdot b}(z)= z/(1+z)^3$ using (\ref{S_chi}). We invert it for $\phi_{a\cdot b}(z)$:
\begin{equation}
z = \frac{\phi_{a\cdot b}(z)}{(1+\phi_{a\cdot b}(z))^3} \ .
\end{equation}
It is a cubic equation for $\phi_{a\cdot b}(z)$. Replacing $\phi_{a\cdot b}(z)$ with the Green's function (\ref{phiG}) we have
\begin{equation}
z^2 G^3_{a \cdot b}(z) - z G_{a \cdot b}(z) + 1 = 0 \ .
\end{equation}
This equation has three solutions given by the Cardano formulas. We select the one which asymptotically behaves as $1/z$ for large $z$. Using (\ref{rhoG}) we find the density
\begin{equation}
\rho_{a \cdot b}(\lambda) = \frac{2^{1/3}3^{1/2}}{12\pi} 
\frac{2^{1/3}\left(27 + \sqrt{27(27-4\lambda)}\right)^{2/3} - 6 
\lambda^{1/3}}{\lambda^{2/3}\left( 27 + 
\sqrt{27(27-4\lambda)}\right)^{1/3}}
\label{WW}
\end{equation} 
for $\lambda \in \left[0,\frac{27}{4}\right]$ and zero otherwise.

We close this section with a few general comments. The multiplication law (\ref{Sab}) or
(\ref{Rprod}) can be derived using solely perturbation expansion for invariant matrix models in the limit $N\rightarrow \infty$ without referring to free probability. We refer the interested reader to \cite{bjn}.  The second comment concerns the multiplication law for matrices which are not positive semi-definite. In the discussion above we restricted the class of matrices to positive semi-definite ones because we wanted to use $\sqrt{a}$ in the definition of the product $a \cdot b$ (\ref{cprod}). For (invariant) Hermitian matrices the product $a\cdot b$ is (invariant) Hermitian. One can also consider the multiplication law for the standard product $ab$ of Hermitian matrices. In this case $a$ and $b$ do not have to be positive semi-definite. The matrix $ab$ is non-Hermitian and thus has complex eigenvalues. Its trace moments $\frac{1}N {\rm Tr} (ab)^k$ are, however, real. Since the S transform depends only on the moments, all  calculations are identical as discussed before for $a\cdot b$ except that the final result $\rho_{ab}(\lambda)$ cannot be interpreted as an eigenvalue density function, but merely as a function to calculate the trace moments of the matrix $ab$:
\begin{equation}
m_{ab,n} = \left\langle \frac{1}{N} {\rm Tr} (ab)^n \right\rangle = 
\int \int d\mu(a) d\mu(b) \frac{1}{N} {\rm Tr} (ab)^n = \int_R d\lambda \rho_{ab}(\lambda) \lambda^n \ .
\end{equation}  
The eigenvalue density has in this case generically a two-dimensional support on the complex plane. For example, the product of two independent Hermitian matrices from the GUE ensemble is a non-Hermitian matrix. Each GUE matrix has real eigenvalues with the eigenvalue distribution given by the Wigner semicircle law. The R transform $R_a(z)=R_b(z)= z$ (we consider standardized GUE with the mean zero and unit variance). Using the R transform form of the multiplication law (\ref{Rprod}) one finds that $R_{ab}(z)=0$ and thus $G_{ab}(z)=1/z$. This means that all the moments $m_{ab,n}=0$. It does not mean, however, that eigenvalue density of the product is zero. To calculate the eigenvalue density in this case one has to use methods that allow one to treat non-Hermitian matrices. We discuss them in the subsequent sections. 
Here we only mention that the eigenvalue density for the product is
\begin{equation}
\rho_{ab}(z) = \left\{\begin{array}{rl} \frac{1}{2\pi} \frac{1}{|z|}  \ , & |z|\le 1 \\
0 \ , & |z| > 1 \end{array} \right.
\label{HH}
\end{equation}
where $z$ is a complex variable. This result was first derived in \cite{bjw} 
with the help of field theoretical methods. Note that the result is invariant 
under rotations around the origin of the complex plane. 

\section{Product of isotropic random matrices}
\label{sec-iso}

The extension from Hermitian to non-Hermitian random matrices is similar to the extension from real to complex random variables in classical probability. In this section we employ this similarity to define isotropic random matrices \cite{bns} being probably the simplest extension of invariant random matrices to non-Hermitian ones. 

Consider a complex random variable with a probability measure 
that is invariant under rotation around the origin of the complex plane $\mu(z) = \mu(z e^{i\alpha})$. The probability density is a function of the distance from the origin
$d\mu(z) = d^2 z \rho\left(|z|\right)$. Such a random variable $z = r e^{i\phi}$ can be constructed as a product of a real non-negative random variable $r$ and a uniformly distributed random phase $\phi\in [0,2\pi)$. When one knows the probability density function for the radial variable $r$, one can easily derive the probability density for $z$ on the complex plane. 

Isotropic random matrices are constructed in a way that imitates this construction. A random matrix $A$ is called isotropic if the probability measure $\mu(A)$ is  invariant under the left and right multiplication by unitary matrices $U,V$: 
$\mu(A)=\mu(U^\dagger A V)$. In Math and Engineering community such random matrices 
are called Bi-Unitarily Invariant (BUI) and in general they can be rectangular.
Here we concentrate on square matrices and use the name isotropic 
in reference to them.

The simplest way of constructing isotropic random matrices is to build them from real semi-positive diagonal random matrices $a$ and Haar unitary matrices $u,v$ 
independent of $a$ (free of $a$ for $N\rightarrow \infty$) 
\begin{equation}
A=v^\dagger au \ .
\end{equation}
Clearly, the diagonal matrix $a$ corresponds to the radial variable $r$ of $z=re^{i\phi}$ and the Haar unitary matrices $v$ and $u$ to the uniform phase $\phi$. There are two of them to ensure the invariance under multiplication by a unitary matrix on both sides. Alternatively one could consider a one-sided invariance, in which case the last equation would take the form $A=au$. Most of the results given below hold in either case. 

Another way to construct such matrices is to define them by a probability measure of the form \cite{fz}
\begin{equation}
d\mu(A) = e^{-N {\rm Tr} V(A^\dagger A)} dA
\label{isoV}
\end{equation}
where $dA = \prod_{ij} d {\rm Im} A_{ij} d{\rm Re} A_{ij}$ is flat measure for $N\times N$ complex matrices and $V(x)$ is a polynomial or a power series in $x$.
There are many other ways of defining isotropic matrices. An interesting property of isotropic random matrices is that independently of the way they are defined, in the limit $N\rightarrow \infty$ there is a one-to-one correspondence between the eigenvalue density of the matrix $A$ and the eigenvalue density of the Hermitian matrix $A^\dagger A$ associated with $A$ \cite{hl}. By construction the matrix $A^\dagger A$ is an invariant positive semi-definite random matrix. We denote the square root of this matrix by $a=\sqrt{A^\dagger A}$. Note that the eigenvalues of $a$ correspond to singular values of $A$. We can use the whole arsenal of methods discussed in the previous section to calculate the eigenvalue distribution of such matrices or their products. 

The correspondence between the eigenvalue distribution of a large isotropic matrix and the distribution of its singular values was found by Feinberg and Zee \cite{fz} who applied field theoretical perturbative methods to solve the matrix model (\ref{isoV}) and independently 
by Haagerup and Larsen \cite{hl} who used methods of free probability. Below we quote the result in the form given by Haagerup and Larsen \cite{hl}. In the terminology of free probability isotropic matrices in the limit $N\rightarrow \infty$ correspond to variables that are called R-diagonal \cite{ns2}. Here we prefer to call them large isotropic matrices. The invariance of the measure implies that the eigenvalue density of an isotropic matrix is invariant under rotation around the origin of the complex plane. For an isotropic matrix $A$ it is convenient to define radial cumulative density function as the probability that a random eigenvalue of the isotropic matrix lies on the complex plane within the distance $x$ from the origin: $F_A(x) ={\rm Prob}(|\lambda_A|\le x)$. The eigenvalue density is related to the radial cumulative function as
\begin{equation}
\rho_A(z) = \frac{1}{2\pi|z|} F'_A(|z|) + p_0 \delta^{(2)}(z)\ , 
\label{rhoF}
\end{equation}
where $p_0 = F_A(0)$ is a possible point mass located at the origin.
It corresponds to the probability of finding zero eigenvalues in the 
eigenvalue spectrum of $A$. For $p_0>0$ the number of zero eigenvalues
is a finite fraction of all eigenvalues.
 
Knowing $F_A(x)$ one can calculate the density $\rho_A(z)$ and vice versa. With the help of this function one can write an equation which relates the eigenvalue distributions of $A$ and $a=\sqrt{A^\dagger A}$ \cite{hl}
\begin{equation}
\label{HLtheorem}
S_{a^2}\left(F_A(x)-1\right)=\frac{1}{x^{2}} \ , 
\end{equation}
for $x>0$. This equation does not fix the value of the radial cumulative
function for $x=0$. It is fixed by an additional equation
\begin{equation}
\label{HL2} 
p_0=F_A(0)=F_a(0) \ ,
\end{equation}
where $F_a(x)=\int_0^x \rho_a(\lambda) d\lambda$ 
is the cumulative eigenvalue density function for $a$.
This equation merely means that the number of zero eigenvalues of
$A$ is equal to the number of zero eigenvalues of $a$.
The value $F_a(0)$ is generically zero except the 
situation when there is a point measure (Dirac delta) at zero
in the eigenvalue density $\rho_a(\lambda)$, as for instance
in the distribution 
\begin{equation}
\rho_a(\lambda) = (1-r) \delta(0) + 
\frac{1}{2\pi \lambda} \sqrt{(\lambda_+-\lambda)(\lambda-\lambda_-)} \ ,
\end{equation}
with $\lambda_\pm = (1 \pm \sqrt{r})^2$, $0<r<1$ and the support being
$\{0\} \cup [\lambda_-,\lambda_+]$. This distribution naturally arises
in the context of random matrices and it is called
anti-Wishart distribution in physics literature and free 
Poissonian in math and engineering literature. The first term on the right hand side 
means that the fraction of zero eigenvalues of the corresponding random matrix $a$
is $(1-r)$. The corresponding isotropic random matrix 
$A=au$ `inherits' all zero eigenvalues from $a$ and thus 
the corresponding fraction of zero eigenvalues of $A$ is also $1-r$. 
   
Equations (\ref{HLtheorem}) and (\ref{HL2}) 
provide an implicit relation between $\rho_A(z)$ and $\rho_a(\lambda)$ 
that allows one to calculate one from another. 

A few comments are in order.
Equation (\ref{HLtheorem}) was first published in physical literature 
in a slightly different (less transparent) albeit equivalent form \cite{fz}. 
In this paper it was also observed that for $N\rightarrow \infty$ 
the support of the eigenvalue density of an isotropic random matrix has 
the shape of a single ring on the complex plane and that the radii of the ring 
can be derived from the eigenvalue density of $a$. The external radius is 
equal to $R_{e}=(\int_0^\infty \rho_{a}(\lambda) \lambda^2 d\lambda)^{1/2}$ 
and the internal one to 
$R_{i}=(\int_0^\infty \rho_{a}(\lambda) \lambda^{-2} d\lambda)^{-1/2}$. 
The ring reduces to a disk when $R_i=0$. In the paper \cite{hl} the discussion
was formalized and all statements were proven rigorously using concepts of free probability. In addition to that the S transform form of equation 
(\ref{HLtheorem}) and the discussion of the point measure for zero eigenvalues 
$p_0\ne 0$ (\ref{HL2}) were given. From the discussion \cite{hl} it follows that
in general there are three different types of solutions for $F_A$. 
The ring solution that corresponds to $F_A(x)$ growing monotonically 
from $0$ to $1$ for $x \in [R_i,R_e]$ and for $R_i>0$, 
the disk solution that corresponds to $F_A(x)$ growing monotonically 
from $0$ to $1$ for $x \in [0,R_e]$ and the disk solution with 
a finite fraction of zero eigenvalues $p_0>0$ that corresponds to 
$F_A(x)$ growing monotonically from $p_0$ to $1$ for $x \in [0,R_e]$.
 
Let us illustrate how the Haagerup-Larsen theorem (\ref{HLtheorem})
works in practice by giving an example. Consider a random matrix $A$ from the Ginibre ensemble \cite{g,ks}. It is an $N \times N$ random matrix whose elements are 
independent identically distributed, normally distributed, centered complex random variables. The probability measure for $A$ is
\begin{equation}
d\mu(A) \propto e^{-N {\rm Tr} A^\dagger A} dA
\end{equation}
where $dA$ is flat measure for $N\times N$ complex matrices. Obviously $A$ is an isotropic random matrix. The limiting eigenvalue density for this matrix for $N\rightarrow \infty$ is given by the uniform distribution on the unit disk \cite{g,ks}
\begin{equation}
\label{Ginibre}
\rho_A(z) = \frac{1}{\pi} \ , \ |z|\le 1 \ .
\end{equation}
It is zero outside the disk. The cumulative density $F_A(x) = \int_0^x dr 2\pi r \rho_A(r)$ is
\begin{equation}
F_{A}(x) = x^2 \ , \ x\le 1 \ .
\label{FG}
\end{equation}
Inserting it into equation (\ref{HLtheorem}) we obtain an equation for the S transform 
\begin{equation}
\label{Sa2}
S_{a^2}(z) = \frac{1}{1+z} \ .
\end{equation}
Next we find (\ref{S_chi}) $\chi_{a^2}(z) = \frac{z}{(z+1)^2}$,
the moment generating function (the inverse function for $\chi_{a^2}$)
and the Green's function (\ref{phiG}) 
\begin{equation}
G_{a^2}(z) = \frac{1}{2} - \frac{1}{2} \sqrt{\frac{z-4}{z}} \ .
\end{equation}
Finally we calculate the eigenvalue density of $a^2$ (\ref{rhoG})
\begin{equation}
\label{a2}
\rho_{a^2}(\lambda) = \frac{1}{2\pi} \sqrt{\frac{4-\lambda}{\lambda}} \ , \
\lambda \in [0,4] \ ,
\end{equation}
and correspondingly of $a$
\begin{equation}
\label{a}
\rho_a(\lambda) = \frac{1}{\pi} \sqrt{4-\lambda^2} \ , \ \lambda \in [0,2] \ .
\end{equation}
This is the density of singular values of the Ginibre matrix. We see that it is given by the quarter-circle law. 

Let us now discuss the product of $n$ independent isotropic matrices $A_1$, $A_2$, $\ldots$, $A_n$
\begin{equation}
P = A_1 A_2 \ldots A_n 
\end{equation}
where $A_i$'s are independent isotropic matrices \cite{bjw,bls}. The goal is to calculate 
the eigenvalue density and the singular value density in the limit $N\rightarrow \infty$
having the knowledge of eigenvalue densities for the matrices in the product.
Since the product itself is an isotropic matrix we can use equation (\ref{HLtheorem})
to write
\begin{equation}
\label{HLP}
S_{p^2}\left(F_P(x)-1\right)=\frac{1}{x^{2}} \ . 
\end{equation}
The S transform for $p^2=PP^\dagger$ can be calculated
using the multiplication law (\ref{Sab})
\begin{equation}
S_{p^2}(z) = S_{a_1^2 \cdot a_2^2 \cdot \ldots \cdot a_n^2}(z) = \prod_{i=1}^n S_{a_i^2}(z) \ .
\label{paaa}
\end{equation}
Since the right-hand side of this equation is a product of complex numbers, it does not depend on the order of multiplication. If instead of $P$ we considered a product of matrices permuted with a permutation $\pi$ 
\begin{equation}
P_\pi = A_{\pi(1)} A_{\pi(2)} \ldots A_{\pi(n)} \ ,
\end{equation}
we would obtain the S transform $S_{p^2_\pi}(z)$ identical to $S_{p^2}(z)$ because  multiplication on the right hand side of (\ref{paaa}) is commutative. Therefore $F_P(x) = F_{P_\pi}(x)$ are equal 
for any permutation $\pi$. In other words, the product of isotropic matrices is spectrally commutative in the limit $N\rightarrow \infty$. 

Isotropic random matrices possess a striking property that has no counterpart in classical probability. Consider the product $P=A_1 \ldots  A_n$ of independent identically distributed isotropic random matrices. This can be realized by picking up $n$ independent matrices from the same ensemble and multiplying them. It turns out that in the limit $N\rightarrow \infty$ such a product has exactly the same eigenvalue density as the $n$-th power $Q=A^n$ of one matrix picked up at random from this ensemble \cite{bns}. Actually 
it was already observed in \cite{hl} that singular values of $Q$ and $P$ are identically distributed. This observation can be trivially extended to eigenvalues
if one additionally assumes that $p_0=0$ (\ref{HL2}). 
This is a sort of spectral self-averaging of multiplication in the limit $N\rightarrow \infty$ because a single matrix from the ensemble is sufficiently representative in the $n$-fold product to give the same result as $n$ independent random realizations. 
The proof is short so we recall it here in the form given in \cite{bns} 
where it was tacitly assumed that $p_0=0$. All constituents needed for this  
proof can be found already in \cite{hl}. 

We assume that $p_0=F_A(0)=F_a(0)=0$ (\ref{HL2}). 
The S transform for the product of independent identically distributed matrices is the $n$-th power of the S transform for a single one $S_{p^2}(z) = S^n_{a^2}(x)$ as follows from equation (\ref{paaa}) in which all $S_{a^2_i}(z)$ are replaced by $S_{a^2}(z)$. Inserting
$S^n_{a^2}(x)$ in place of $S_{p^2}(z)$ in equation (\ref{HLP}) and 
taking the $n$-th root of both sides we get
\begin{equation}
S_{a^2}\left(F_P(x)-1\right)=\frac{1}{x^{2/n}} \ .
\end{equation}
Comparing the last equation to equation (\ref{HLtheorem}) we see that $F_P(x) = F_A(x^{1/n})$. We can now argue that the radial cumulative density $F_Q(x)$
for the power $Q$ is given by the same expression. 
Indeed, using directly the definition of radial 
cumulative density we have 
\begin{equation}
F_Q(x) = {\rm Prob}(|\lambda_Q|\le x) = {\rm Prob}(|\lambda_A|^n\le x) = 
 {\rm Prob}(|\lambda_A| \le x^{1/n}) =F_A(x^{1/n})
\end{equation}
for $x>0$, and for $F_Q(x)=F_P(x)=0$ for $x=0$ and hence $F_P(x)=F_Q(x)=F_A\left(x^{1/n}\right)$. This completes the proof.
If $F_A(0)=p_0 > 0$ then the proof breaks down at $x=0$. Recall that $p_0$
is the probability of drawing at random a zero-eigenvalue from the spectrum of 
random matrix $A$.
The corresponding probability for the product of $n$ independent $A$ matrices 
is $F_P(0) = 1 - (1-p_0)^n$, while for the $n$-th power 
$F_Q(0) = 1 - p_0$, so we see that $F_P(0) \ne F_Q(0)$ for $p_0>0$.
 
In particular, one can use the product-power equivalence to calculate the eigenvalue density of the product $P=A_1\ldots A_n$ of $n$ independent Ginibre matrices 
in the limit $N\rightarrow \infty$. We know that $F_A(x)=x^2$ for $x\in [0,1]$ for one Ginibre matrix (\ref{FG}) thus for the product of $n$
\begin{equation}
F_P(x) = x^\frac{2}{n} \ , \ x \in [0,1]  \ .
\label{GGG}
\end{equation}
The eigenvalue density for the product is (\ref{rhoF}) 
\begin{equation}
\rho_P(z) = \frac{1}{n\pi} |z|^{\frac{2(1-n)}{n}} \ , \ |z|\le 1
\end{equation}
and zero outside the unit circle. This calculation reproduces in a very simple way the result that was originally obtained by different methods \cite{bjw}. This result can be generalized also to products of rectangular Gaussian matrices \cite{bjlns} 
or products  \cite{agt,os} of Wigner matrices \cite{w} with non-Gaussian entries.

One can also find the distribution of singular values for the product of Ginibre matrices \cite{cnz,bls}. Inserting radial cumulative density function (\ref{GGG}) to the Haagerup-Larsen formula (\ref{HLtheorem}) one can find the Green's function for $p^2=P^\dagger P$ \cite{bls}
\begin{equation}
\label{Gn}
z^n G^{n+1}_{p^2}(z) = z G_{p^2}(z) - 1 \ 
\end{equation}
and calculate the density function $\rho_{p^2}(\lambda)$ by using (\ref{rhoG}). For $n=1$ it gives the Wishart distribution (\ref{Wish}); for $n=2$ the distribution given in equation (\ref{WW}). One can find an explicit expression for any $n$ \cite{pz}.
The last step is to change variables $\lambda \rightarrow \lambda^2$ to go from $\rho_{p^2}(\lambda)$ to $\rho_p(\lambda)$. 

To summarize, large isotropic random matrices are exceptional in many respects. There exists a one-to-one correspondence between the densities of eigenvalues and singular values \cite{hl}. Multiplication of independent isotropic matrices is spectrally commutative and self-averaging in the large $N$ limit \cite{bns,bls}. One can also show that if the eigenvalue density for an isotropic matrix has a power-law singularity at zero $\rho_A(z) \sim |z|^{-s}$ with $0<s<2$, then the density of singular values behaves at zero as $\rho_p(\lambda) \sim \lambda^{-\frac{s}{4-s}}$ \cite{bls}. 

\section{General multiplication law}
\label{sec-nh}

So far we have discussed invariant and isotropic random matrices. We are now going to discuss the more general case of products of non-Hermitian matrices. The multiplication law for non-Hermitian matrices was derived in reference \cite{bjn} by field theoretical methods.

To build some intuition let us discuss Gaussian non-Hermitian random matrices. One can construct such matrices from independent identically distributed Hermitian Gaussian matrices from the GUE ensemble. The probability measure for GUE matrices is $d\mu(a) \sim e^{-N {\rm Tr}  a^2/2} da$, where $da$ is flat measure for Hermitian matrices. 
Let us take two independent such matrices $a$ and $b$. The following combination
\begin{equation}
X= \frac{1}{\sqrt{2}}\left(a + ib\right)
\end{equation}
is a complex random matrix; $a$ gives rise to Hermitian sector of $X$ and $ib$ to the anti-Hermitian one. The resulting matrix $X$ is a Ginibre random matrix \cite{g,ks} with the uniform eigenvalue distribution on the unit disk (\ref{Ginibre}). One can also define elliptic random matrices \cite{ks,gi} 
\begin{equation}
X= \cos(\alpha) a + i \sin(\alpha) b
\end{equation}
being linear combinations of Hermitian and anti-Hermitian random matrices with different coefficients controlled by a mixing parameter $\alpha$. For $\alpha=0$ the random matrix $X$ reduces to the Hermitian matrix $a$, for $\alpha=\pi/4$ to the Ginibre matrix and for $\alpha=\pi/2$ to the anti-Hermitian matrix $ib$. The eigenvalue density of the matrix $X$ is constant on the support being an ellipse whose oblateness depends on $\alpha$.
The probability measure for matrix $X$ can be calculated from the measure $d\mu(a,b)= d\mu(a) d\mu(b) \propto e^{-N {\rm Tr} (a^2+b^2)/2} da db$ that factorizes since $a$ and $b$ are independent. Changing variables to $X=\cos(\alpha) a + i \sin(\alpha) b$ and $X^\dagger = \cos(\alpha) a - i \sin(\alpha) b$ one finds the probability measure for $X$ \cite{fks,scss}  
\begin{equation}
d\mu(X) \propto \exp\left( -N \frac{1}{1-\tau^2} {\rm Tr} \left( XX^\dagger - \
\tau (XX + X^\dagger X^\dagger)\right)\right) dX \ ,
\label{muX}
\end{equation}
where $\tau=\cos(2\alpha)$. For this measure one can calculate two-point correlation functions $\left\langle X_{ab} X_{cd} \right\rangle = \int d\mu(X) X_{ab} X_{cd}$. There are four such functions~:
\begin{equation}
\begin{split}
\left\langle X^\dagger_{ab} X^\dagger_{cd} \right\rangle = \tau \frac{1}{N} \delta_{ad}\delta_{bc} \ , \quad & 
\left\langle X^\dagger_{ab} X_{cd} \right\rangle = \frac{1}{N} \delta_{ad}\delta_{bc} \ ,\\
\left\langle X_{ab} X^\dagger_{cd} \right\rangle = \frac{1}{N} \delta_{ad}\delta_{bc} 
\ , \quad &
\left\langle X_{ab} X_{cd} \right\rangle = \tau \frac{1}{N} \delta_{ad}\delta_{bc} \ ,
\end{split}
\end{equation}
that can be compactly written in the matrix form 
\begin{equation}
\left(\begin{array}{rr}
\left\langle X^\dagger_{ab} X^\dagger_{cd} \right\rangle  & 
\left\langle X^\dagger_{ab} X_{cd} \right\rangle \\
\left\langle X_{ab} X^\dagger_{cd} \right\rangle  &
\left\langle X_{ab} X_{cd} \right\rangle \end{array}\right)=
\left(\begin{array}{rr}
\tau & 1 \\
1 & \tau \end{array}\right) \otimes \frac{1}{N} \delta_{ad}\delta_{bc} \ .
\label{tau}
\end{equation}
The matrix on the right hand side of the last equation defines the correlation structure of the model \cite{bjn}. It plays an important role, as we shall see below. The parameter $\tau$ takes values between $-1$ and $1$. For Ginibre matrix it is $0$, and for GUE it is $1$.

Without going into details we quote the law of non-Hermitian matrix multiplication and refer the interested reader to \cite{bjn}. We present it in the R transform form (\ref{Rprod}). The main difference as compared to the Hermitian case (\ref{Rprod}) is that the R transform is not a complex function but a quaternionic one. We denote quaternions by calligraphic letters. The R transform maps quaternions ${\cal G}$ onto quaternions ${\cal G} \rightarrow {\cal R}(\cal G)$. The multiplication law reads 
\begin{equation}
\begin{split}
{\cal R}_{AB}({\cal G}_{AB})  &=            \left[{\cal R}_A({\cal G}_B)\right]^L 
                                      \left[{\cal R}_B({\cal G}_A)\right]^R \ , \\
\left[{\cal G}_A\right]^R  &= {\cal G}_{AB} \left[{\cal R}_A({\cal G}_B)\right]^L \ , \\
\left[{\cal G}_B\right]^L  &=          \left[{\cal R}_A({\cal G}_B)\right]^R {\cal G}_{AB} \ .
\end{split}
\label{RAB}
\end{equation}
Here ${\cal G}_A = {\cal G}_A(z)$, ${\cal G}_B = {\cal G}_B(z)$ and
${\cal G}_{AB} = {\cal G}_{AB}(z)$ correspond to generalized quaternionic Green's functions for $A$, $B$ and $AB$. They have the structure:
\begin{equation}
{\cal G}(z) = \left( \begin{array}{rr} 
g(z) & i b(z) \\ i \bar{b}(z) & \bar{g}(z) \end{array} \right) \ ,
\label{qGz}
\end{equation}
where $b(z)$ and $g(z)$ are complex functions. We suppressed the indices of the corresponding matrix  ($A$, $B$ or $AB$) in the last equation. The density is related to the upper left element $g(z)$ of the generalized quaternionic Green's function 
\begin{equation}
\rho(z) = \frac{1}{\pi} \partial_{\bar{z}} g(z) \ .
\label{elec}
\end{equation}
The superscripts $L$ and $R$ refer to the left and right adjoint actions 
of the following unitary matrix 
\begin{equation}
U=\left(\begin{array}{cc} \left(\frac{z}{\bar{z}}\right)^{\frac{1}{8}} & 0 \\ 0 & \left(\frac{\bar{z}}{\bar{z}}\right)^{\frac{1}{8}} \end{array}\right) 
\label{ULR}
\end{equation} 
on matrices in the square brackets 
\begin{equation} 
[X]^L = UXU^\dagger \ , \quad  [X]^R = U^\dagger X U  \ .
\label{lr}
\end{equation} 
The diagonal elements of $U$ are constructed from $z$ being the argument of the Green's functions (\ref{qGz}). The generalized R transform is related to the Green's function by an equation analogous to (\ref{gr})
\begin{equation}
{\cal G}(z) = \left({\cal Z} - {\cal R}({\cal G}(z))\right)^{-1}
\label{GR}
\end{equation}
where 
\begin{equation}
{\cal Z} = \left(\begin{array}{rr} z & 0 \\ 0 & \bar{z} \end{array}\right) . 
\end{equation}
This law looks complicated but in practice it reduces to a set of algebraic equations for $g(z)$ from which one can calculate the eigenvalue density by taking the derivative with respect to $\bar{z}$ (\ref{elec}). Before we give an example let us mention an electrostatic analogy \cite{scss}. In this analogy eigenvalues of a random matrix represent positions of charges in two-dimensions in the given potential, the function $g(z)$ is interpreted as the electric field $\vec{g}=(g_x,g_y)$ with components $g_x={\rm Re} g(z)$ and $g_y=- {\rm Im}g(z)$ and equation (\ref{elec}) as the Gauss law in two dimensions \cite{scss}.
It takes a more familiar form when written in the components of $z=x+iy$:
\begin{equation}
\rho(z) = \frac{1}{2\pi} (\partial_x g_x + \partial_y g_y) = \frac{1}{2\pi} 
\vec{\nabla} \cdot \vec{g} \ .
\end{equation}
We now illustrate how to use the multiplication law (\ref{RAB}) to calculate the eigenvalue density of the product. We consider as an example the multiplication of Gaussian matrices constructed from the centered elliptic matrix $X$ (\ref{muX}) by shifting and rescaling
\begin{equation}
A = q \mathbbm{1} + \sigma X \ .
\label{Aqs}
\end{equation}
The scale parameter $\sigma$ is a real number and the shift parameter $q$ is a complex number. The quaternionic R transform for $A$ is 
\begin{equation}
{\cal R}_A\left(\left( \begin{array}{rr} g & i b \\ i \bar{b} & \bar{g} \end{array} \right)\right) = 
\left( \begin{array}{rr} q & 0 \\ 0 & \bar{q} \end{array} \right) + 
\sigma \left( \begin{array}{rr} \tau g & i b \\ i \bar{b} & \tau \bar{g} \end{array} \right) \ . 
\label{Rqs}
\end{equation}
There are only two terms on the right-hand side because the matrix is Gaussian and thus  all higher order cumulants vanish (\ref{R}). The second term on the right-hand side is directly related to the matrix on the right hand side of equation (\ref{tau}) that defines the two-point correlation structure of the model \cite{bjn,bjw}. 

\begin{figure}
\begin{center}
\includegraphics[width=0.5\textwidth]{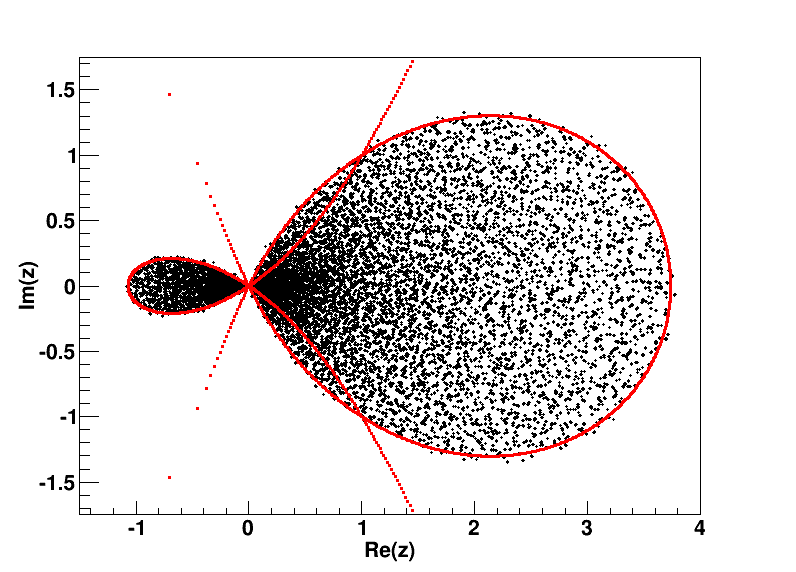}
\end{center}
\caption{Solid line represents the contour of the eigenvalue density for the product $(\mathbbm{1} + X)(\mathbbm{1} + H)$ where $X$ is a standardized Ginibre matrix and $H$ is an independent standardized Hermitian GUE matrix. The curves forming the contour are calculated analytically \cite{bs} using the multiplication law (\ref{RAB}). The extrapolation of these curves outside the support is shown in red dots. Black dots correspond to eigenvalues collected for ten matrices of size $1000 \times 1000$ generated by Monte-Carlo simulations. \label{1X1H} }
\end{figure}

The simplest example is $q=0$, $\sigma=1$ and $\tau=0$ (\ref{Aqs}). It corresponds to the Ginibre matrix. Let us calculate the eigenvalue density for the product of two such matrices. The first equation in the multiplication law (\ref{RAB}) gives
\begin{equation}
{\cal R}_{AB}({\cal G}_{AB})  =   \left( \begin{array}{cc} 0 & i c b_B \\ i \bar{c} \bar{b}_B & 0 \end{array} \right) \left( \begin{array}{cc} 0 & i \bar{c} b_A \\ i c \bar{b}_A & 0 \end{array} \right) = \left( \begin{array}{cc} -c^2 b_B \bar{b}_A & 0 \\
0 &  -\bar{c}^2  \bar{b}_B b_A \end{array} \right) \ .
\end{equation}
We used a shorthand notation $c= (z/\bar{z})^{1/4}$. Because $A$ and $B$ are identically distributed, we search a symmetric solution $b_A(z)=b_B(z)$ and $g_A(z)=g_B(z)$. 
\begin{equation}
{\cal R}_{AB}({\cal G}_{AB})  =  \left( \begin{array}{cc} -c^2 |b_A|^2 & 0 \\
0 &  -\bar{c}^2  |b_A|^2 \end{array} \right) \ .
\label{Raa}
\end{equation}
Now we can use the two remaining equations (\ref{RAB}). Because of the symmetry it is enough to use only one of them, for example the first one. Replacing 
${\cal G}^{-1}_{AB}$ by ${\cal Z} - {\cal R}_{AB}({\cal G}_{AB})$ (\ref{GR}) we 
can write this equation as
\begin{equation}
\left({\cal Z} - {\cal R}_{AB}({\cal G}_{AB})\right) \left[{\cal G}_A\right]^R  = \left[{\cal R}_A({\cal G}_B)\right]^L \ .
\end{equation} 
We can now eliminate ${\cal R}_{AB}$ in the last equation using (\ref{Raa}). 
Writing also $\left[{\cal G}_A\right]^R$ and $\left[{\cal R}_A({\cal G}_B)\right]^L$
in the explicit matrix form we have
\begin{equation}
\left( \begin{array}{cc} z+c^2 |b_A|^2 & 0 \\  
0 & \bar{z} +\bar{c}^2 |b_A|^2 \end{array} \right)
\left( \begin{array}{cc} g_A & i \bar{c} b_A \\  
i c \bar{b}_A & \bar{g}_A \end{array} \right) = \left( \begin{array}{cc} 0 & i c b_A \\  
i \bar{c} \bar{b}_A & 0 \end{array} \right)  \ ,
\end{equation}
where $c= (z/\bar{z})^{1/4}$, as before. It is a set of equations for $g_A=g_A(z)$ and $b_A=b_A(z)$. It has a trivial solution $g_A(z)=b_A(z)=0$ and a non-trivial one $g_A(z)=0$ and 
$|b_A(z)|^2 = 1-|z|$. The two solutions coincide at the unit circle $|z|=1$.
The trivial solution holds for $|z|>1$ and the nontrivial one for $|z| \le 1$. Note that for both of them $g_A(z)=0$. Inserting these solutions to (\ref{Raa}) we obtain ${\cal R}_{AB}$ and then using (\ref{GR}) we derive the final result for the generalized Green's function of the product
\begin{equation}
{\cal G}_{AB}(z) = \left(\begin{array}{rr} \sqrt{\frac{\bar{z}}{z}} & 0 \\
0 & \sqrt{\frac{z}{\bar{z}}} \end{array} \right) 
\end{equation}
inside the unit circle $|z|\le 1$, and 
\begin{equation}
{\cal G}_{AB}(z) = \left(\begin{array}{rr} \frac{1}{z} & 0 \\
0 & \frac{1}{\bar{z}} \end{array} \right) 
\end{equation}
outside. Applying the Gauss law (\ref{elec}) to the upper left element of ${\cal G}_{AB}(z)$ (denoted by $g_{AB}(z)$) we find the density 
\begin{equation}
\rho_{AB}(z) = \frac{1}{\pi} \partial_{\bar{z}} g_{AB}(z) = \frac{1}{2\pi|z|} 
\label{rab}
\end{equation}
for $|z|\le 1$, and $\rho_{AB}(z)=0$ otherwise. The calculations were done for $\tau_A=\tau_B=0$. One, however, (quite surprisingly) obtains  the same result for the product of matrices with arbitrary $\tau_A$ and $\tau_B$ \cite{bjw}. The reason for this is the following. Looking at the R transform (\ref{Rqs}) we see that $\tau$ appears only in the combination $\tau g$ (or, writing it separately for $A$ and $B$, in combinations $\tau_A g_A$ and $\tau_B g_B$). The dependence on $\tau$'s disappears for $g_A(z)=g_B(z)=0$. Since in the solution discussed above we had $g_A(z)=g_B(z)=0$, we conclude that this solution is independent of $\tau_A$ and $\tau_B$ and thus it holds for any $\tau_A$ and $\tau_B$. In particular, the product $AB$ of two Hermitian GUE matrices ($\tau_A=\tau_B=1$) or any elliptic random matrices has exactly the same eigenvalue density (\ref{rab}), which is spherically symmetric.
\begin{figure}
\begin{center}
(a)
\includegraphics[width=0.35\textwidth]{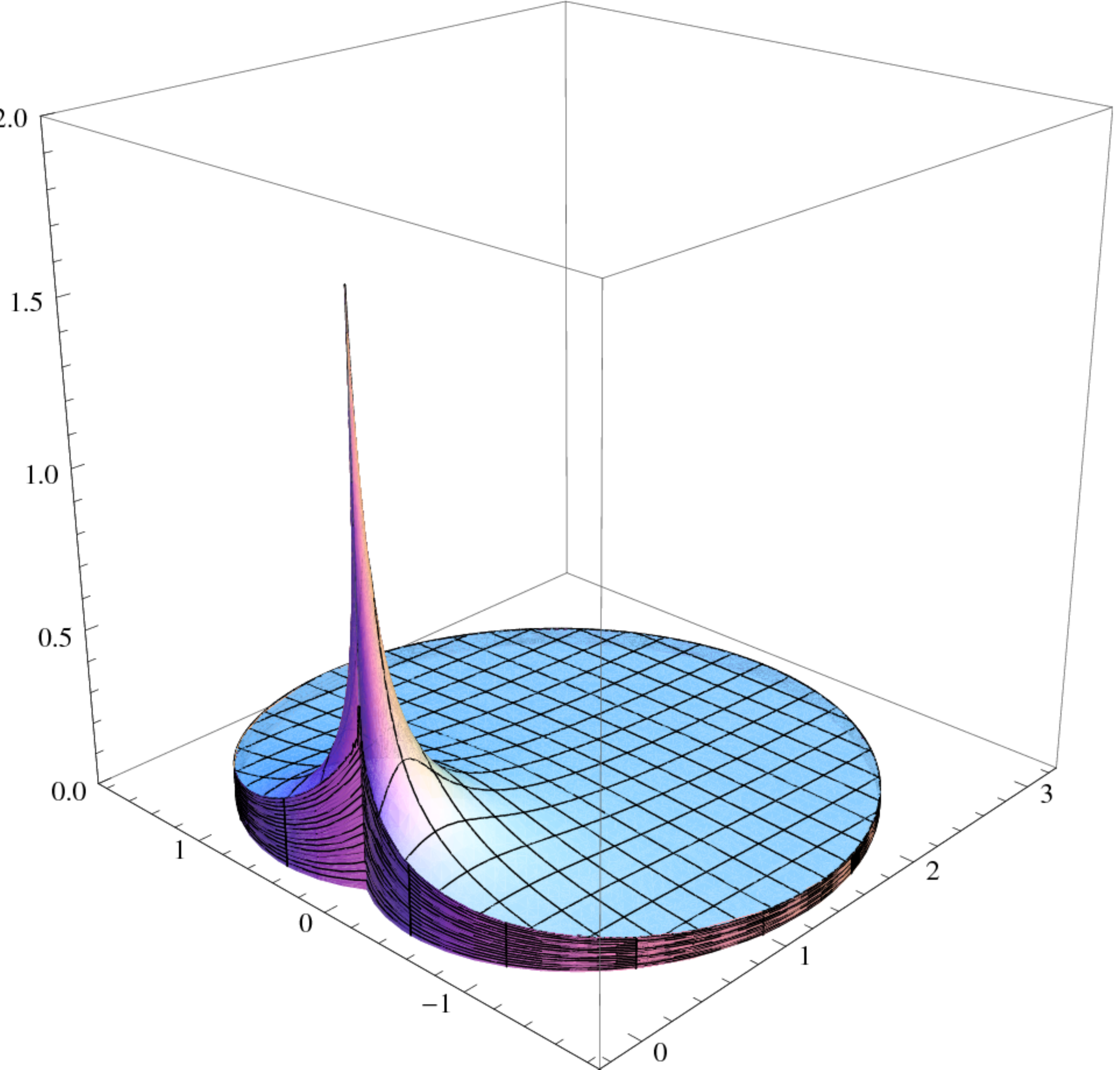} \quad 
(b) \includegraphics[width=0.5\textwidth]{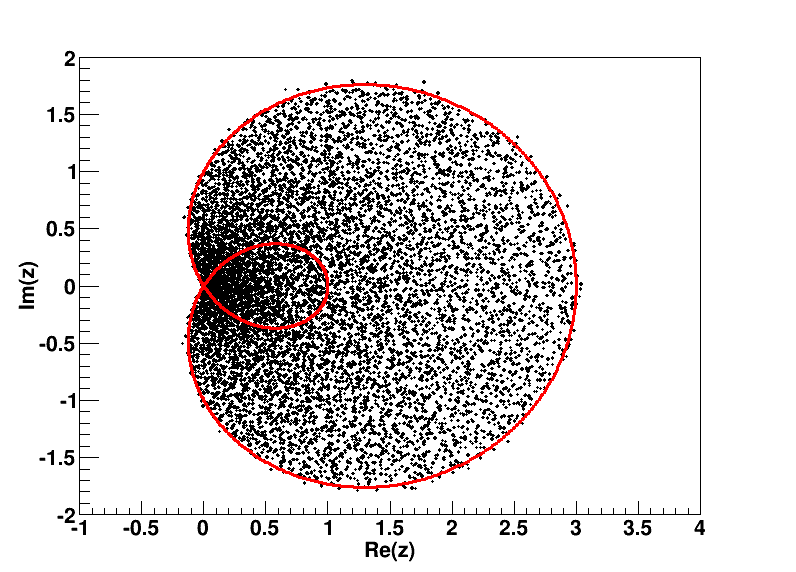}
\end{center}
\caption{(a) Eigenvalue density for the product $(\mathbbm{1} + X_1)(\mathbbm{1} + X_2)$ where $X_1$ and $X_2$ are independent standardized Ginibre matrices, each having the eigenvalue density uniformly distributed on the unit circle. The result is derived analytically \cite{bjn}. The horizontal axes correspond to the real and imaginary axes of the complex plane and the vertical one to the value of the eigenvalue density $\rho(z)$. The eigenvalue density has a support surrounded by a contour given by the Pascal's lima\c{c}on. 
(b) Solid line represents the Pascal's lima\c{c}on. Dots correspond to eigenvalues collected for ten matrices of size $1000 \times 1000$ generated by Monte-Carlo simulations. A few eigenvalues lie outside the limiting contour but this effect is expected by finite size analysis. The number of outliers decreases when the matrix size increases.
\label{1X1X} }
\end{figure}
In a similar way one can find the eigenvalue density for the product of matrices with non-zero constants $q$ (\ref{Aqs}) but the calculations are tedious. In figure \ref{1X1X}.(a) we show an analytic result for the eigenvalue density for $(\mathbbm{1} + X_1)(\mathbbm{1} + X_2)$ where $X_1$ and $X_2$ are independent Ginibre matrices \cite{bjn}. The eigenvalue support is surrounded by a contour $r=1+2\cos \phi$ that corresponds to the the Pascal's lima\c{c}on. In figure \ref{1X1X}.(b) we compare this theoretical prediction to Monte Carlo simulations of large, finite matrices. The agreement is very good.  As another example, in figure \ref{1X1H} we show a comparison between Monte Carlo results and the analytic prediction for the contour of the support for the eigenvalue density of the product $(\mathbbm{1} + H)(\mathbbm{1} + X)$, where $H$ is the GUE Hermitian matrix and $X$ is the Ginibre matrix. Again the comparison shows a good agreement. The details of the calculations and other examples can be found in \cite{bs}. 
We summarize this section shortly by emphasizing that there exists an algorithm to calculate eigenvalue densities for products of non-Hermitian matrices. This algorithm works for large random matrices ($N\rightarrow \infty$) with probability measures of the type $d\mu (X) \propto e^{- N {\rm Tr} V(X,X^\dagger)} dX$ with a self-adjoint potential $\left(V(X,X^\dagger)\right)^\dagger = V(X,X^\dagger)$ which is bounded from below \cite{bjn}.    
\section{Conclusions}

\label{sec-concl}

Using free probability one can find an explicit formula for the moment generating function for the product of invariant random Hermitian matrices in the limit $N\rightarrow \infty$. This formula allows to calculate trace moments $\left\langle \frac{1}N {\rm Tr} (ab)^k \right\rangle$ for independent (free) random matrices $a,b$ and to determine the eigenvalue density of the matrix $\sqrt{a} b \sqrt{a}$ (if $a$ is positive semi-definite) from eigenvalue densities of $a$ and $b$. We discussed how to extend the formalism to non-Hermitian matrices to calculate the eigenvalue density of the product $ab$, which is generically a non-Hermitian matrix. The extension has been worked out using field theoretical methods of planar graphs enumeration. The multiplication law (\ref{RAB}) is conveniently written in the R transform form (\ref{Rprod}). The R transform is a quaternionic function. One can use this law to find eigenvalue densities for products of large non-Hermitian random matrices.

Isotropic matrices are a special class of non-Hermitian random matrices. These matrices have two exceptional features as far as matrix multiplication is concerned. Isotropic random matrices are spectrally commutative and self-averaging in the limit $N\rightarrow \infty$. The first notion means that the eigenvalue distribution of the product of such matrices does not depend on the order of matrix multiplication. For example, the matrix $abcd$ has the same eigenvalue density as $bdac$. The second one means that the product of $n$ independent isotropic identically distributed matrices has the same eigenvalue distribution as the $n$-th power of a single matrix.

We finish by mentioning that recently a progress has been also made in the understanding of microscopic properties of eigenvalues statistics for products of random matrices. For example, an explicit form of the joint probability distribution for eigenvalues and singular values of the product of Ginibre matrices has been explicitly found \cite{ab,akw}. Using these results one can determine not only the eigenvalue density but also joint eigenvalue distribution and eigenvalue correlations for products of large random matrices. One can also infer finite size corrections to the limiting laws discussed in this paper. 

\subsection*{Acknowledgments}
The author would like to thank Romuald Janik, Andrzej Jarosz, Giacomo Livan, Maciej A. Nowak, Gabor Papp, Artur Swiech and Ismail Zahed for many interesting discussions and a fruitful collaboration on the subject and Jeremi Ochab for critical reading of the
manuscript. The author also wants to thank the organizers of the conference ``Inference, Computation, and Spin Glasses'' held in Sapporo, July 28th--30th 2013, for a kind invitation to give a talk on this conference. The research presented in the paper was financed by Grant No. DEC-2011/02/A/ST1/00119 of National Centre of Science in Poland. The participation in the conference was supported by the grant: Grant-in-Aid for Scientific Research on Innovative Areas, MEXT, Japan.

\end{document}